\documentclass[a4paper]{jpconf}
\usepackage{graphicx}
\begin{document}
\title{Review of the theoretical heavy-ion physics}

\author{E. L. Bratkovskaya$^{1,3}$, W. Cassing$^2$, P. Moreau$^3$, T. Song$^2$}

\address{$^1$ GSI Helmholtzzentrum f\"{u}r Schwerionenforschung GmbH,  Darmstadt, Germany}
\address{$^2$ Institute for Theoretical Physics, University of Giessen, Giessen, Germany}
\address{$^3$ Institute for Theoretical Physics, University of Frankfurt, Frankfurt, Germany}

\ead{E.Bratkovskaya@gsi.de}

\begin{abstract}
In this contribution we  briefly give an overview of the theoretical
models used to describe experimental data from heavy-ion collisions
from $\sqrt{s_{NN}} \approx $ 4 GeV to ultra-relativistic energies
of  $\sqrt{s_{NN}} \approx $ 5 TeV.  We highlight the successes and
problems of statistical or hadron-resonance gas models, address the
results of macroscopic approaches like hydrodynamics (in different
hybrid combinations) as well as the results from microscopic
transport approaches in comparison to experimental data. Finally,
the transport coefficients like shear $\eta$ and bulk viscosity
$\zeta$  - entering the
macroscopic models - are confronted with results from lattice QCD in
thermal equilibrium for vanishing chemical potential.
\end{abstract}

\section{Introduction}
The dynamics of the early universe in terms of the 'Big Bang' may be
studied experimentally by relativistic nucleus-nucleus collisions
from Alternating Gradient Synchrotron (AGS) to Large-Hadron-Collider
(LHC) energies in terms of 'tiny bangs' in the laboratory. With
sufficiently strong parton interactions, the medium in the collision
zone can be expected to achieve local equili\-brium after some
initial delay and exhibit approximately hydrodynamic
flow~\cite{Ol92,HK02,Sh09}. In these collisions a new state of
strongly interacting matter is created, being characterized by a
very low shear viscosity $\eta$ to entropy density $s$ ratio,
$\eta/s$, close to a nearly perfect fluid~\cite{Sh05,GMcL05}.
Lattice QCD (lQCD) calculations~\cite{Cheng08,aori10} indicate that
a crossover region between hadron and quark-gluon matter should have
been reached in these experiments (at least at higher energies).
Apart from a deconfinement transition also a restoration of chiral
symmetry should occur at about the same critical temperature in case
of vanishing baryon chemical potential. Whereas at low chemical
potential the transition is known to be a crossover \cite{L1,L2} it
is presently unclear if there will be a critical point in the QCD
phase diagram marking the transition to a first-order domain
\cite{Xu}. Furthermore, it is doubted that the restoration of chiral
symmetry and the deconfinement transition will happen at the same
point in the phase diagram once high baryon chemical potentials are
encountered \cite{Rob1,Rob2,sasaki}. This situation will be met
experimentally in heavy-ion collisions at FAIR/NICA energies in the
future \cite{CBMbook}.

Since the hot and dense matter produced in relativistic heavy-ion
collisions appears only for a couple of fm/c, it is a challenge for
experiment to investigate its properties. The differential spectra
of hadrons with light quarks/antiquarks provide information about
the bulk dynamics whereas abundances and differential spectra of
hadrons with strange/antistrange quarks shed light on the chemical
equilibration processes. Furthermore, the electromagnetic emissivity
of the matter produced in heavy-ion collisions is tested by direct
photon spectra as well as dileptons which might provide additional
information on the properties of vector resonances in a dense
hadronic medium. Also the heavy flavor mesons are considered to be
promising probes since the production of heavy flavor requires a
large energy-momentum transfer and takes place early in the
heavy-ion collisions, and - due to the large energy-momentum
transfer - should be described by perturbative quantum
chromodynamics (pQCD). The produced heavy flavor then interacts with
the hot dense matter  (of partonic or hadronic nature) by exchanging
energy and momentum which is controlled by a spatial diffusion
constant $D_s(T,\mu_B)$. Let's have a brief look at the various
model concepts.

\section{Statistical, macroscopic and microscopic models}

\subsection{Hadron resonance gas (HRG) or statistical models}
In case of very strong interactions of the degrees of freedom in the
collision zone of relativistic heavy-ion collisions one might infer
that a thermal and chemical equilibrium has been achieved (at least
at freezeout) and the final hadronic spectra can be described by a
grand-canonical ensemble assuming the conservation of energy,
particle number and volume on average. When looking at particle
ratios the volume drops out - implying similar freezeout conditions
and collective flow for all hadrons - and one is left with
essentially two Lagrange parameters that are attributed to a
temperature $T$ and  baryon chemical potential $\mu_B$. Thus -
looking at central collisions of Au+Au (Pb+Pb) - one can extract a
freezeout line in the QCD phase diagram by fitting the measured
particle ratios (dominantly at midrapidity) at different bombarding
energies. In fact, the results of such fits are in a good agreement
with experimental observation from AGS to top LHC energies
\cite{Andro,Stachelnew} over many orders of magnitude once some
parameter for the excluded volume of hadrons ($r_0 \approx$ 0.3 fm)
is choosen properly in order to reduce the net density. Especially
at top LHC energies only a single parameter $T$ survives since
$\mu_B \approx$ 0. In principle the inclusion of resonances is akin
to an interacting theory of 'fundamental' hadrons with attractive
interactions; the concept of an excluded volume decreases the
density and increases the pressure thus simulating additional
repulsive interactions. These models are reminiscent of Van der
Waals gases and recent statistical models actually are formulated
along this line or incorporate explicit baryon-baryon interactions
\cite{Goren1,Goren2,Goren3}. Accordingly, the formulation of
interacting hadron resonance gas models (IHRG) will provide a link
between the actual interaction parameters and fundamental many-body
theories (e.g. Brueckner) or $S$-matrix approaches.

However, the question about the dynamics of equilibration and the
generation of collective flow in heavy-ion collisions remains open
in the HRG or IHRG approaches. Furthermore, the evaluation of photon
or dilepton spectra cannot be addressed in the grandcanonical (or
canonical) models since the real and virtual photons are not in
equilibrium with their environment due to the low electromagnetic
coupling ($\alpha_e \approx 1/137$). Some of these questions,
however, can be addressed in macroscopic models.

\subsection{Hydro and hybrid models}
In order to obtain some information on the space-time dynamics of
heavy-ion collisions one often employs hydrodynamical models which
are of one-fluid \cite{HK02} or three-fluid \cite{Ivanov} nature. In
the one-fluid models the initial conditions for the hydro-evolution
(at some finite time $t_0$ assumed for local equilibration) are
essentially fixed by the final hadron spectra and the generation of
collective flow in the fluid follows from the local pressure
gradients (adopting some equation of state (EoS) for the fluid). In
this case one can model various EoS relating to a hadronic one, to
lattice QCD or a model EoS with a first order phase transition in
order to test the sensitivity of observables like collective flow
coefficients $v_n$ ($n$=1,2,3,4,..) as a function of bombarding
energy and centrality of the collision. However, in ideal
hydrodynamical simulations it was found that the elliptic flow as a
function of transverse momentum $v_2(p_T)$ was overestimated in
comparison to experimental data at RHIC thus signalling a finite
shear viscosity $\eta$. Actually, there is a lower limit on the
ratio of shear viscosity over entropy density $\eta/s \ge 1/(4\pi)$
\cite{Kovtun:2004de} such that viscous hydrodynamics had to be
employed \cite{Hydrorev}. Furthermore, also a finite (and even
large) bulk viscosity $\zeta(T)$ for temperatures close to $T_c
\approx$ 158 MeV should be incorporated. With the appearance of two
additional transport coefficients $\eta(T)$ and $\zeta(T)$ one had
to specify their functional form since results from pQCD turned out
to be fully misleading. On the other hand a significant triangular
flow $v_3(p_T)$ demonstrated the importance of initial-state
fluctuations that had to be incorporated in the initial conditions
of the hydro phase \cite{triangular}. Furthermore, resonant hadronic
scattering - after chemical freezeout - had to be included since the
hadrons still keep interacting after chemical freezeout
\cite{Aichelin17}. This lead to the development of hybrid models
which incorporate three different type of model components:
\begin{itemize}
\item{i) the initial nonequilibrium phase to specify the initial state fluctuations or initial flow}
\item{ii) viscous hydro for the partonic (fluid) phase}
\item{iii) hadronic 'afterburner' for resonant interactions in the hadronic phase after freezeout.}
\end{itemize}
Due to the matching of the different phases a couple of new
parameters enter such models that define the matching conditions.
Accordingly, a multi-parameter approach (on the scale of $\sim$ 15
independent parameters) emerges that has to be optimized in
comparison to a multitude of experimental data in order to extract
physical information on the transport coefficients. This has been
done within a Bayesian analysis by a couple of authors and some
proper information could be extracted so far on $\eta/s(T)$ as well
as for the charm diffusion coefficient $D_s(T)$
\cite{Bass0,Bass1,Bass2}. For explicit results we refer the reader
to Refs. \cite{Bass0,Bass1,Bass2,Bayesian,Pratt1,Pratt2,pratt3}. We
note in passing that within such approaches semi-central and central
nucleus-nucleus collisions at ultra-relativistic energies can well
be described \cite{ulli} but an application to elementary
high-energy $p+p$ or $\pi+p$ reactions is difficult/questionable.

In this class of models we mention also the ultra-relativistic
quantum molecular dynamics (UrQMD) hybrid approach which starts with
UrQMD \cite{UrQMD1} for the initial nonequilibrium phase on an event
by event basis, switches to hydro after approximate equilibration in
local cells of higher energy density, continues with a hydro
evolution until freezeout (at equal times) and follows with UrQMD to
describe the final hadronic rescatterings \cite{Petersen}. By
construction such hybrid models may be used for lower (AGS) energies
as well as for ultra-relativistic (LHC) energies. A systematic study
of transport properties in the fluid phase is still not available so
far. Further attempts incorporating a color glass condensate (CGC) for
the initial conditions, IP-glasma or EPOS2 initial conditions
\cite{Schenke1,Schenke2,Glasma,Ipglasma,EPOS1,EPOS2,EPOS3} provide also a good
description of the collective flows as a function of bombarding
energy and collision centrality at RHIC and LHC energies.

A further advantage of hydro or hybrid models is that one can
calculate the differential photon and dilepton production by
integration of microscopic production rates in space and time
\cite{Rapp,Gale}. Especially in the partonic phase the AMY rates
\cite{AMY} are employed for photon production by most of the authors
whereas the evolution of the electromagnetic emissivity in the
hadronic phase differs substantially within the different variants.

\subsection{Microscopic transport models}
Whereas early microscopic transport models have been developed for
the dynamics of hadrons employing a nuclear matter EoS and cross
sections based on experimental data or effective hadronic
Lagrangians \cite{BUU,IQMD,GiBUU} later versions have included the
formation and decay of strings \cite{UrQMD1,HSD,Ehehalt} to
incorporate multi-particle production with increasing energy which
becomes essential at AGS and SPS energies. However, when applied to
heavy-ion collisions at RHIC energies a couple of problems emerged
since a number of observables (elliptic flow of charged hadrons,
transverse mass spectra of hadrons, intermediate mass dileptons
etc.) could no longer be properly described by  hadron-string
degrees of freedom \cite{Brat04,BratPRL}. This lead to the
formulation of transport models for partonic degrees of freedom of
Boltzmann-type that were first based on pQCD based cross sections
\cite{AMPT,BAMPS} with an ideal gas EoS for the partonic phase.
Since pQCD scattering cross sections between massless partons turned
out too low in order to describe the elliptic flow of hadrons
measured experimentally, either effective (enhanced) two-body cross
sections have been used \cite{AMPT2} or additional $2
\leftrightarrow 3$ channels have been added as in BAMPS
\cite{BAMPS}. The formation of hadrons is usually performed by
coalescence either in momentum space or - more recently - in phase
space. Another branch of transport models is based on NJL-like
approaches including a coupling to a scalar mean field and/or a
vector mean field \cite{Greco1,Rudy}. In these models the partons
have a finite dynamical mass and the binary cross sections are
either extracted from the NJL Lagrangian \cite{Rudy} or
parameterized to simulate a finite $\eta/s$ (as in hydro models)
\cite{Greco12}. All these approaches provide a reasonable
description of experimental data at RHIC energies as well as for LHC
energies.

The Parton-Hadron-String Dynamics (PHSD) transport
approach~\cite{CB09,PHSDreview} is a microscopic covariant dynamical
model for strongly interacting systems formulated on the basis of
Kadanoff-Baym equations \cite{Cassing:2008nn} for Green's functions
in phase-space representation (in first order gradient expansion
beyond the quasiparticle approximation). The approach consistently
describes the full evolution of a relativistic heavy-ion collision
from the initial hard scatterings and string formation through the
dynamical deconfinement phase transition to the strongly-interacting
quark-gluon plasma (sQGP) as well as hadronization and the
subsequent interactions in the expanding hadronic phase as in the
Hadron-String-Dynamics (HSD) transport approach \cite{HSD}. The
transport theoretical description of quarks and gluons in the PHSD
is based on the Dynamical Quasi-Particle Model (DQPM) for partons
that is constructed to reproduce lQCD results for a quark-gluon
plasma in thermodynamic equilibrium~\cite{Cassing:2008nn} on the
basis of effective propagators for quarks and gluons. The DQPM is
thermodynamically consistent and the effective parton propagators
incorporate finite masses (scalar mean-fields) for gluons/quarks as
well as a finite width that describes the medium dependent reaction
rate. For fixed thermodynamic temperature $T$ the partonic width's
$\Gamma_i(T)$ fix the effective two-body interactions that are
presently implemented in the PHSD~\cite{Vitalii}. The PHSD differs
from conventional Boltzmann approaches in a couple of essential
aspects:
\begin{itemize}
\item{it incorporates dynamical quasi-particles due to the finite
width of the spectral functions (imaginary part of the propagators)
in line with complex retarded selfenergies;} \item{it involves
scalar mean-fields that substantially drive the collective flow in
the partonic phase;} \item{it is based on a realistic equation of
state from lattice QCD and thus reproduces the speed of sound
$c_s(T)$ reliably;} \item{the hadronization is described by the
fusion of off-shell partons to off-shell hadronic states (resonances
or strings);} \item{all conservation laws (energy-momentum, flavor
currents etc.) are fulfilled in the hadronization contrary to
coalescence models;} \item{the effective partonic cross sections no
longer are given by pQCD and are 'defined' by the DQPM in a
consistent fashion. These cross sections are probed by transport coefficients
(correlators) in thermodynamic equilibrium by performing PHSD
calculations in a finite box with periodic boundary conditions
(shear- and bulk viscosity, electric conductivity, magnetic
susceptibility etc. \cite{Vitaly2,Ca13}).}
\end{itemize}
The transition from the partonic to hadronic degrees-of-freedom (for
light quarks/antiquarks) is described by covariant transition rates
for the fusion of quark-antiquark pairs to mesonic resonances or
three quarks (antiquarks) to baryonic states, i.e. by the dynamical
hadronization \cite{CB09}.  Note that due to the off-shell nature of
both partons and hadrons, the hadronization process described above
obeys all conservation laws (i.e. four-momentum conservation and
flavor current conservation) in each event, the detailed balance
relations and the increase in the total entropy $S$. In the hadronic
phase PHSD is equivalent to the hadron-strings dynamics (HSD) model
\cite{HSD} that has been employed in the past from SIS to SPS
energies. On the other hand the PHSD approach has been tested for
p+p, p+A and relativistic heavy-ion collisions from lower SPS to LHC
energies and been successful in describing a large number of
experimental data including single-particle spectra, collective flow
\cite{PHSDreview} as well as electromagnetic probes \cite{photons}
or charm observables \cite{Song:2015sfa,Song:2015ykw}.

Apart from deconfinement the chiral symmetry restoration (CSR)
addresses another aspect of the QCD phase diagram in the ($T,
\mu_B$)-plane as an additional  transition between a phase with
broken and a phase with restored chiral symmetry. As in case of the
QCD deconfinement phase transition, the boundaries of the CSR phase
transition line are not well known. Lattice QCD (lQCD) calculations
show that at vanishing baryon chemical potential $\mu_B$=0 the CSR
takes place at roughly the same critical temperature and energy
density as the deconfinement phase transition which is a crossover.
At finite baryon chemical potential lQCD calculations cannot be
performed due to the sign problem and one must rely on effective
models (or extrapolations) in order to study the QCD phase
transitions \cite{CBMbook}. Different models support the idea that
at finite chemical potential a partially restored phase is achieved
before the deconfinement occurs \cite{Rob1,Rob2}. In order to
distinguish the two phases of such a transition, effective models
use the scalar quark condensate $\langle \bar q q \rangle$ as an
order parameter. As the baryon density and temperature increase, the
scalar quark condensate $\langle \bar q q \rangle$ is expected to
decrease from a non-vanishing value in the vacuum to $\langle \bar q
q \rangle\approx\,0$ which corresponds to CSR. Since $\langle \bar q
q \rangle$ is not a measurable quantity, it is crucial to determine
experimental observables which are sensitive to this quantity. Since
long the dilepton spectroscopy has been in the focus in this respect
since in a chirally restored phase the spectral functions of the the
$\rho$- and the $a_1$-meson should become identical. However, no
clear evidence has been achieved so far \cite{dileptons}. On the
other hand, the enhanced strangeness production at AGS and lower SPS
energies was found to be a signature of CSR \cite{PHSD_CSR,Alessia}
within the PHSD approach where the local scalar quark condensate was
evaluated along the line of the Hellmann-Feynman theorem from the
scalar density of hadrons (cf. Ref. \cite{Alessia} for details).

Accordingly, microscopic transport approaches provide a bridge from
$p+p$, to $p+A$ and $A+A$ collisions and allow for a transparent
interpretation of differential particle spectra, collective flow and
electromagnetic observables from experimental studies at various
facilities and a wide energy range. Open problems are still
many-body reactions - except for 2 $\leftrightarrow$ 3 channels
\cite{Cas2002,Eduard} - and the dynamical modeling of first-order
transitions in transport. The formation of clusters is still a task
to be solved as well as the inclusion of chiral anomalies.

\section{Transport coefficients}
Information on the QCD phase diagram from strongly interacting matter
does not only come from experimental studies but can also be
addressed by {\it ab initio} QCD calculations on a discrete
(Euklidean) space-time lattice. Due to the Fermion-sign problem
direct lQCD calculations cannot be performed at finite  chemical
potential, however, valuable information can be inferred from lQCD
calculations at imaginary chemical potentials as well as by Taylor
expansions. Here the second order expansion coefficients - related
to susceptibilities as e.g. $\chi_B = \partial^2 P/\partial \mu_B^2$
- can be evaluated at vanishing $\mu_B$ and provide a first glance
in $\mu_B$ direction at finite temperature $T$. Here $P$ denotes the
pressure which is identical to the (negative) grand-canonical
partition function. Apart from susceptibilities $\chi_x$ also
transport coefficients (shear viscosity $\eta(T)$, bulk viscosity
$\zeta(T)$, electric conductivity $\sigma_e(T)$, spatial diffusion
constant $D_s(T)$ for charm quarks etc.) can be calculated on the
lattice although with still quite some uncertainties. These
transport coefficients either enter the viscous hydro calculations
(Section 2.2) as input or can be confronted with the Bayesian
results from hydro (or hybrid) calculations in comparison to a large
set of different observables (cf. Section 2.2).

On the other hand, the microscopic transport models can be studied
also in a finite box at some initial energy density $\epsilon$ and
net-particle number density $n_B$ employing periodic boundary
conditions. Within the Kubo formalism \cite{Kubo} or the
relaxation-time approximation (RTA) \cite{Kapusta} then the
transport coefficients can be determined in equilibrium (after some
finite equilibration time to determine the thermodynamic variables
$T$ and $\mu_B$) and be confronted with results from lQCD. Since in
leading order the relation between pressure $P$ and energy density
$\epsilon$ is relevant or in particular the speed of sound squared
$c_s^2(T)=\partial P/\partial \epsilon$, the microscopic transport
models, that claim to describe experimental data, also have to
reproduce $c_s^2(T)$ to provide a consistent picture. This excludes
those models with massless weakly interacting partons since the EoS
cannot be reproduced in the vicinity of the critical temperature
$T_c$. We note in passing that  explicit comparisons of both methods
(Kubo and RTA) in Ref. \cite{Vitalii} for $\eta/s$ have shown that
the solutions are rather close. This holds especially for the case
of the scattering of massive partons where the transport cross
section is not very different from the total cross section as also
pointed out in Ref. \cite{Plumari:2012ep}.

\begin{figure}[tbh!]
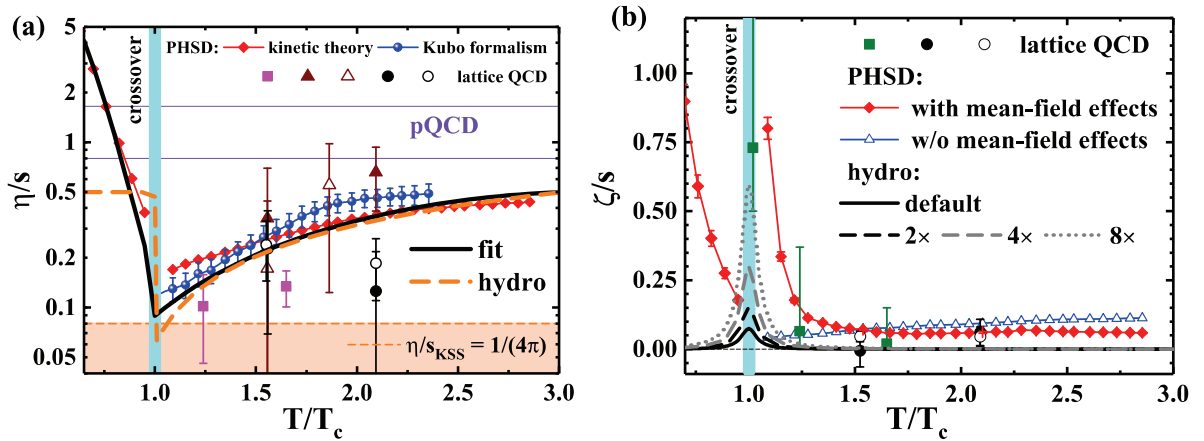
 
\includegraphics[width=14pc,angle=270]{Fig1.eps}
\hspace*{-0.5cm}\includegraphics[width=14pc,angle=270]{Fig2.eps}
\caption{ $\eta/s$ (a) and $\zeta/s$ versus scaled temperature
$T/T_c$.(a) The symbols indicate the PHSD results of $\eta/s$ from
Ref. \cite{Vitalii}, calculated using different methods: the
relaxation-time approximation (red line + diamonds) and the Kubo
formalism (blue line + dots); the black line corresponds to the
parametrization of the PHSD results for $\eta/s$. The orange short
dashed line demonstrates the Kovtun-Son-Starinets bound
\cite{Kovtun:2004de} $( \eta/s )_{KSS } = 1 / (4\pi)$. The orange
dashed line is the $\eta/s$ of the VISHNU hydrodynamical model that
was recently determined by a Bayesian analysis. (b) $\zeta/s$ from
PHSD simulations from Ref. \cite{Vitalii} and the $\zeta/s$ adapted
in the hydrodynamical simulations of Ref. \cite{Basslena}.  The
symbols with (large) error bars are lQCD results from different
groups. The figures are taken from Ref. \cite{Basslena}.}\label{Fig}
\end{figure}

In Fig. \ref{Fig} we display different results for $\eta/s$ (a) and
$\zeta/s$ versus the scaled temperature $T/T_c$. All variants
suggest that $\eta/s$ has a minimum close to $T_c$ whereas $\zeta/s$
shows a maximum close to $T_c$. It is worth noting that especially
for the shear viscosity the results from PHSD simulations from the
relaxation-time approximation (red line + diamonds) and the Kubo
formalism (blue line + dots) are in close agreement with those from
the Bayesian analysis within the VISHNU hydrodynamical model (orange
dashed line) as well as with the results from lQCD. This
demonstrates that the different theoretical methods outlined above
come to approximately the same answers.

\section{Summary}
In this contribution we have briefly discussed the various models
used for the description of observables from relativistic heavy-ion
collisions in the energy range from the AGS to the LHC and pointed
out their successes, range of applications and problems. Whereas
statistical models provide no dynamical information the hydro or
hybrid models need external information with respect to the
transport coefficients and initial conditions/fluctuations. These
models succeed in describing various phenomena of relativistic
heavy-ion collisions and a Bayesian analysis of a large set of
experimental data allows to pin down constraints on the transport
coefficients of interest. On the other hand, microscopic transport
models  provide a bridge from $p+p$, to $p+A$ and $A+A$ collisions
and allow for a transparent interpretation of differential particle
spectra, collective flow and electromagnetic observables from
experimental studies at various facilities and a wide energy range.
It is interesting to note that different methods have almost
converged to the same results for the shear viscosity (cf. Fig. 1)
which demonstrates that complementary strategies lead to a closer
physical understanding of the strongly interacting matter produced
in heavy-ion reactions.  Open problems in microscopic transport
approaches are still many-body reactions,  the dynamical modeling of
first-order transitions, the formation of clusters  as well as the
inclusion of chiral anomalies.

\section*{Acknowledgements}
\label{acknowledgment}
The authors acknowledge valuable discussions with J. Aichelin, S. A. Bass, O.
Linnyk, V. Ozvenchuk,  A. Palmese, E. Seifert,  T. Steinert, V. Toneev,
V. Voronyuk, and Y. Xu. This
work has been supported by the ``HIC for FAIR'' framework of the
``LOEWE'' program, BMBF and DAAD.

\end{document}